# Exact Solutions for Free Vibration Analysis of Thick Laminated Annular/Circular Plates Using Third-Order Shear Deformation Plate Theory

Reza Serajian, Said Asadi

**Abstract.** In this paper, an exact analytical solution for free vibration analysis of thick laminated annular/circular plates is presented. The equilibrium equations are derived according to the Reddy's third order shear deformation plate theory (TSDT). Governing equations are simplified into decoupled equations and solved analytically for plates with different boundary conditions. Then, natural frequency is obtained and the effects of geometrical parameters on annular/circular plates are studied. The accuracy and convergence of the presented method are demonstrated through numerical examples. For free vibration problems, the influences of various thickness ratios, radius ratios, the material property graded indexes and circumferential wave numbers on the lowest non-dimensional frequency are investigated under different circular boundary conditions. In-plane and out-plane mode shapes are presented to illustrate flexural motions of the plate.

## 1. Introduction

Laminated composite structures are among the most important structures widely used in the fields of aerospace, automotive and other engineering industries since they have considerably more stiffness and strength. The behavior of structures composed of several laminates is considerably more complicated than for isotropic ones.

The classical plate theory (CPT) [1] and first order shear deformation theory (FSDT) [2] are commonly used theory for the analysis of laminated composite plates. However, CPT predicts good results for thin plates only, because, the transverse shear deformation is omitted in CPT. FSDT does not satisfy shear stress free conditions at top and bottom surfaces of plates. Further, FSDT is not capable of properly constraining all the displacements at the clamped supports of beams and plates. Higher order shear deformation theories are therefore developed to overcome these limitations of classical laminated plate theory and first order shear deformation theories for the better representation of the bending, buckling and vibration of the laminated composite and sandwich plates [3].

Various higher-order shear deformation plate theories were proposed, including the second-order shear deformation formulation of Whitney and Sun [4] and the third-order shear deformation theory of Lo et al. [5] with 11 unknowns; Kant [6] with six unknowns; Reddy [7] with five unknowns and Hanna and Leissa [8] with four unknowns. However, among the aforementioned higher-order theories, the third-order shear deformation theory of Reddy [7] is the most widely adopted model in the study of plates, especially laminated ones, due to its high efficiency and simplicity.

Third order shear deformation theory, which is one of the equivalent single layer theories, is used. This theory is based on the same assumptions as the classical (CLPT) and first order shear deformation plate theories (FSDT), except that the assumption on the straightness and normality of the transverse normal is relaxed [9,10].

Exact natural frequencies of thick multilayered laminated composite plates were presented by Srinivas and Rao [11] and Noor [12]. Exact solution for the free vibration analysis of plates based on third-order shear deformation plate was determined by Hashemi et al. [13]. Hashemi et al. [14] also used differential transformation method to develop a semi-analytical solution for free vibration and modal stress analyses of circular plates resting on two-parameter elastic foundations. Xu [15] presented three-dimensional exact solutions for the free vibration of laminated transversely isotropic circular, annular and sectorial plates with unusual boundary conditions using a new state space technique. Regarding the finite element method (FEM), Swaminathan and Patil [16,17], and Ganapathi and Makhecha [18] have carried out free vibration analysis of laminated composite and sandwich plates based on higher order shear and normal deformation theory using finite element method. Recently, Sayyad and Ghugal [19] presented a comprehensive review study on the free vibration analysis of laminated composite and sandwich plates.

The present paper deals with an exact analytical solution for free vibration analysis of thick laminated annular/circular plates. The equilibrium equations are derived according to the third order shear deformation plate theory (TSDT). The natural frequency is obtained and the effects of geometrical parameters on annular circular plates are studied.

2. **Basic Equations**

2.1. Kinematics

The displacement field given by Reddy's higher-order[20],

$$u_r(r,\theta,z,t) = u + z\psi_r - \frac{4z^3}{3h^2}\left(\psi_r + \frac{\partial w}{\partial r}\right)$$

$$u_\theta(r,\theta,z,t) = v + z\psi_\theta - \frac{4z^3}{3h^2}\left(\psi_\theta + \frac{\partial w}{r\partial\theta}\right)$$

(1)

$$u_z(r,\theta,z,t) = w(r,\theta,t)$$

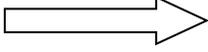

Here u, v and w denote the displacements along the r –and θ- and z directions, respectively, of a point located on the mid-plane of the plate, which coincides with the rθ -plane of the co-ordinate system; $\psi_r, \psi_\theta$ denote the rotations (taken clockwise positive) of a line element, initially perpendicular to the mid-plane, about the θ and r axes, strain-displacement equations of linear elasticity are:

## 2.2. HOOKE'S LAW

The state of stress in the $k_{th}$ layer is given by the Hooke's law as follows [21]:

$$\sigma_{rr}^{(k)} = \frac{E^{(k)}}{1-(v^{(k)})^2}(\varepsilon_{rr} + v^{(k)}\varepsilon_{\theta\theta})$$

$$\sigma_{\theta\theta}^{(k)} = \frac{E^{(k)}}{1-(v^{(k)})^2}(\varepsilon_{\theta\theta} + v^{(k)}\varepsilon_{rr}) \qquad (2)$$

$$\sigma_{r\theta}^{(k)} = \frac{E^{(k)}}{(1+v^{(k)})}\varepsilon_{r\theta}$$

$$\sigma_{rz}^{(k)} = 2G_z^k \varepsilon_{rz}$$

$$\sigma_{\theta z}^{(k)} = 2G_z^k \varepsilon_{\theta z}$$

$$\begin{Bmatrix} \sigma_{rr}^{(k)} \\ \sigma_{\theta\theta}^{(k)} \\ \tau_{\theta z}^{(k)} \\ \tau_{rz}^{(k)} \\ \tau_{r\theta}^{(k)} \end{Bmatrix} = \begin{bmatrix} Q_{11}^{(k)} & Q_{12}^{(k)} & 0 & 0 & 0 \\ Q_{12}^{(k)} & Q_{22}^{(k)} & 0 & 0 & 0 \\ 0 & 0 & Q_{44}^{(k)} & 0 & 0 \\ 0 & 0 & 0 & Q_{55}^{(k)} & 0 \\ 0 & 0 & 0 & 0 & Q_{66}^{(k)} \end{bmatrix} \begin{Bmatrix} \varepsilon_{rr} \\ \varepsilon_{\theta\theta} \\ \varepsilon_{\theta z} \\ \varepsilon_{rz} \\ \varepsilon_{r\theta} \end{Bmatrix}$$

(3)

Where $Q_{ij}^{(k)}$ are the reduced stiffnesses and $k$ is the layer number.

$$Q_{11}^{(k)} = Q_{22}^{(k)} = \frac{E^{(k)}}{1-(v^{(k)})^2} \quad , \quad Q_{12}^{(k)} = Q_{21}^{(k)} = \frac{E^{(k)}}{1-(v^{(k)})^2} v^{(k)} \quad ,$$

$$Q_{44}^{(k)} = Q_{55}^{(k)} = 2G_z^k$$

$$Q_{66}^{(k)} = \frac{E^{(k)}}{(1+v^{(k)})} \tag{4}$$

**3. Equilibrium equations**

The equilibrium equations are derived by using Hamilton's principle [22] The kinetic energy and the strain energy associated with the problem are,

$$T = \iiint_V \rho(\dot{u}_r^2 + \dot{u}^2_\theta + \dot{u}^2_z)dV$$

(5)

$$\delta U = \iiint_V (\sigma_{rr}\delta\varepsilon_{rr} + \sigma_{\theta\theta}\delta\varepsilon_{\theta\theta} + 2\sigma_{r\theta}\delta\varepsilon_{r\theta} + 2\sigma_{\theta z}\delta\varepsilon_{\theta z} + 2\sigma_{rz}\delta\varepsilon_{rz})dV$$

(6)

where $\rho$ is mass of density of the material. The principle can be stated in analytical form as,

$$\int_{t_1}^{t_2} (\delta T - \delta U)dt = 0$$

(7)

where $\delta$ indicates a variation with respect to r, $\theta$ and z.

Substituting stress and strain relations into Eq. (7) and integrating the equation by parts and collecting the coefficients of $\delta u, \delta v, \delta \psi_r, \delta \psi_\theta$ and $\delta w$, the following equations of motion are obtained.

$$\delta u: (rN_{rr})_{,r} + N_{r\theta,\theta} - N_{\theta\theta} = I_{11}\ddot{u} + I_{22}\ddot{\psi}_r - \frac{4}{3h^2}I_{44}\ddot{w}_{,r}$$

$$\delta v: (rN_{r\theta})_{,r} + N_{\theta\theta,\theta} + N_{\theta\theta} = I_{11}\ddot{v} + I_{22}\ddot{\psi}_\theta - \frac{4}{3h^2}\frac{1}{r}I_{44}\ddot{w}_{,\theta}$$

$$\delta \psi_r: M_{rr,r} + \frac{1}{r}M_{r\theta,\theta} + \frac{1}{r}(M_{rr} - M_{\theta\theta}) - Q_r - \frac{4}{3h^2}\left(P_{rr,r} + \frac{1}{r}P_{r\theta,\theta} + \frac{1}{r}(P_{rr} - P_{\theta\theta})\right)$$
$$+ \frac{4}{h^2}R_r$$
$$= I_{22}\ddot{u} + I_{33}\ddot{\psi}_r - I_{55}\ddot{w}_{,r} \quad \delta \psi_\theta: M_{r\theta,r} + \frac{1}{r}M_{\theta\theta,\theta} + \frac{2}{r}M_{r\theta} - Q_\theta$$
$$- \frac{4}{3h^2}\left(P_{r\theta,r} + \frac{1}{r}P_{\theta\theta,\theta} + \frac{1}{r}(P_{rr} - P_{\theta\theta})\right) + \frac{4}{h^2}R_r$$
$$= I_{22}\ddot{v} + I_{33}\ddot{\psi}_\theta - I_{55}\frac{1}{r}\ddot{w}_{,\theta} \quad \delta w: Q_{r,r} + \frac{1}{r}Q_{\theta,\theta} + \frac{1}{r}Q_r$$
$$+ \frac{4}{3h^2}\left(P_{rr,rr} + \frac{2}{r}P_{rr,r} - \frac{1}{r}P_{\theta\theta,r} + \frac{1}{r^2}P_{\theta\theta,\theta\theta} + \frac{2}{r}P_{r\theta,r\theta} + \frac{2}{r^2}P_{r\theta,\theta}\right)$$

(8)

$$-\frac{4}{h^2}\left(R_{r,r} + \frac{1}{r}R_r + \frac{1}{r}R_{\theta,\theta}\right) = I_{11}\ddot{w} + I_{44}(\ddot{u}_{,r} + \frac{1}{r}\ddot{v}_{,\theta}) + I_{55}(\ddot{\psi}_{r,r} + \frac{1}{r}\ddot{\psi}_{\theta,\theta}) - I_{77}\nabla^2\ddot{w}$$

Also, the boundary conditions are obtained as

$$\text{at } r = const. \begin{cases} \text{either } \delta u_r = 0, & \text{or } N_r = 0 \\ \text{either } \delta u_\theta = 0, & \text{or } N_{r\theta} = 0 \\ \text{either } \delta \psi_r = 0, & \text{or } M_r - \frac{4}{3h^2}P_r = 0 \\ \text{either } \delta \psi_\theta = 0, & \text{or } M_{r\theta} - \frac{4}{3h^2}P_{r\theta} = 0 \\ \text{either } \delta w = 0, & \text{or } Q_r + \frac{4}{3h^2}\left(P_{r,r} + \frac{2}{r}P_{r\theta,\theta}\right) - \frac{4}{h^2}R_r + \frac{4}{3h^2}\frac{1}{r}(P_r - P_\theta) = 0 \\ \text{either } \delta w_{,r} = 0, & \text{or } P_r = 0 \end{cases}$$

(9)

$$at\ \theta = const. \begin{cases} either\ \delta u_r = 0, & or\ N_{r\theta} = 0 \\ either\ \delta u_\theta = 0, & or\ N_\theta = 0 \\ either\ \delta\psi_r = 0, & or\ M_{r\theta} - \dfrac{4}{3h^2}P_{r\theta} = 0 \\ either\ \delta\psi_\theta = 0, & or\ M_\theta - \dfrac{4}{3h^2}P_\theta = 0 \\ either\ \delta w = 0, & or\ Q_\theta + \dfrac{4}{3h^2}\left(P_{\theta,\theta} + 2P_{r\theta,r}\right) - \dfrac{4}{h^2}R_\theta + \dfrac{8}{3h^2}\dfrac{1}{r}P_{r\theta} = 0 \\ either\ \delta w_{,r} = 0, & or\ P_\theta = 0 \end{cases} \quad (10)$$

In Eqs. (8), the stress resultants $N$, $M$, $P$, $Q$, $R$ and the inertias $I$ are defined by,

$$(I_1, I_2, I_3, I_4, I_5, I_7) = \int_{-h/2}^{h/2} \rho(1, z, z^2, z^3, z^4, z^6)dz$$

$$\bar{I}_3 = I_3 - \frac{8}{3h^2}I_5 + \frac{16}{3h^4}I_7$$

$$\bar{I}_5 = I_5 - \frac{4}{3h^2}I_7$$

$$(N_{rr}, N_{\theta\theta}, N_{r\theta}) = \int_{-h/2}^{h/2} (\sigma_{rr}, \sigma_{\theta\theta}, \sigma_{r\theta})dz$$

$$(M_{rr}, M_{\theta\theta}, M_{r\theta}) = \int_{-h/2}^{h/2} (\sigma_{rr}, \sigma_{\theta\theta}, \sigma_{r\theta})z\,dz \qquad (11)$$

$$(P_{rr}, P_{\theta\theta}, P_{r\theta}) = \int_{-h/2}^{h/2} (\sigma_{rr}, \sigma_{\theta\theta}, \sigma_{r\theta})z^3 dz$$

$$(Q_r, Q_\theta) = \int_{-h/2}^{h/2} (\sigma_{rz}, \sigma_{\theta z})dz$$

$$(R_r, R_\theta) = \int_{-h/2}^{h/2} (\sigma_{rz}, \sigma_{\theta z})z^2 dz$$

Equations (11) can be simplified for laminated plates as,

$$I_1 = \sum_{k=1}^{N} \rho^k (z_{k+1} - z_k)$$

$$I_2 = \sum_{k=1}^{N} \frac{1}{2}\rho^k (z_{k+1}^2 - z_k^2)$$

$$I_3 = \sum_{k=1}^{N} \frac{1}{3}\rho^k (z_{k+1}^3 - z_k^3)$$

$$I_4 = \sum_{k=1}^{N} \frac{1}{4}\rho^k (z_{k+1}^4 - z_k^4)$$

$$I_5 = \sum_{k=1}^{N} \frac{1}{5}\rho^k (z_{k+1}^5 - z_k^5)$$

$$I_7 = \sum_{k=1}^{N} \frac{1}{7}\rho^k (z_{k+1}^7 - z_k^7)$$

(12)

and,

$$N_{rr} = \sum_{k=1}^{N} \left(\frac{E^k}{1-(\nu^k)^2}\right)\left(\left(u_{,rr} + \nu^k \left(\frac{1}{r}u_{,r} + \frac{1}{r}u_{\theta,\theta}\right)\right)(z_{k+1} - z_k)\right.$$

$$\left. + \frac{1}{2}\left(\psi_{r,r} + \nu^k \left(\frac{1}{r}\psi_r + \frac{1}{r}\psi_{\theta,\theta}\right)\right)(z_{k+1}^2 - z_k^2)\right)$$

$$-\frac{1}{3h^2}\left(\psi_{r,r} + w_{,rr} + \nu^k \left(\frac{1}{r}\psi_r + \frac{1}{r}w_{,r} + \frac{1}{r}\psi_{\theta,\theta} + \frac{1}{r^2}w_{,\theta\theta}\right)\right)(z_{k+1}^4 - z_k^4)\right)$$

$$N_{\theta\theta} = \sum_{k=1}^{N} \frac{E^k}{1-(\nu^k)^2}\left(\left(\frac{1}{r}u_{,r} + \frac{1}{r}u_{\theta,\theta} + \nu^k u_{,rr}\right)(z_{k+1} - z_k)\right.$$

$$\left. + \frac{1}{2}\left(\frac{1}{r}\psi_r + \frac{1}{r}\psi_{\theta,\theta} + \nu^k \psi_{r,r}\right)(z_{k+1}^2 - z_k^2)\right.$$

$$\left. -\frac{1}{3h^2}\left(\frac{1}{r}\psi_r + \nu^k \psi_{r,r} + \frac{1}{r}\psi_{\theta,\theta} + \frac{1}{r}w_{,r} + \nu^k w_{,rr} + \frac{1}{r^2}w_{,\theta\theta}\right)(z_{k+1}^4 - z_k^4)\right)$$

$$N_{r\theta} = \frac{1}{2}\sum_{k=1}^{N} \frac{E^k}{1+\nu^k}\left(\left(\frac{1}{r}u_{r,\theta} - \frac{1}{r}u_\theta + u_{\theta,r}\right)(z_{k+1} - z_k) + \frac{1}{2}\left(\frac{1}{r}\psi_{r,\theta} - \frac{1}{r}\psi_\theta + \psi_{\theta,r}\right)(z_{k+1}^2 - z_k^2)\right.$$

$$\left. -\frac{1}{3h^2}\left(\frac{1}{r}\psi_{r,\theta} - \frac{1}{r}\psi_\theta + \psi_{\theta,r} - \frac{2}{r^2}w_{,\theta} + \frac{2}{r}w_{,r\theta}\right)(z_{k+1}^4 - z_k^4)\right)$$

(13)

and,

$$M_{rr} = \sum_{k=1}^{N} \frac{E^k}{1-(\nu^k)^2}\left(\frac{1}{2}\left(u_{r,r} + \nu^k\left(\frac{1}{r}u_r + \frac{1}{r}u_{\theta,\theta}\right)\right)(z_{k+1}^2 - z_k^2) + \frac{1}{3}\left(\psi_{r,r} + \nu^k\left(\frac{1}{r}\psi_r + \frac{1}{r}\psi_{\theta,\theta}\right)\right)(z_{k+1}^3 - z_k^3)\right.$$

$$\left. -\frac{4}{15h^2}\left(\psi_{r,r} + w_{,rr} + \nu^k\left(\frac{1}{r}\psi_r + \frac{1}{r}\psi_{\theta,\theta} + \frac{1}{r}w_{,r} + \frac{1}{r^2}w_{,\theta\theta}\right)\right)(z_{k+1}^5 - z_k^5)\right)$$

$$M_{\theta\theta} = \sum_{k=1}^{N} \frac{E^k}{1-(\nu^k)^2}\left(\frac{1}{2}\left(\frac{1}{r}u_r + \nu^k u_{r,r} + \frac{1}{r}u_{\theta,\theta}\right)(z_{k+1}^2 - z_k^2) + \frac{1}{3}\left(\frac{1}{r}\psi_r + \nu^k\psi_{r,r} + \frac{1}{r}\psi_{\theta,\theta}\right)(z_{k+1}^3 - z_k^3)\right.$$

$$\left. -\frac{4}{15h^2}\left(\frac{1}{r}\psi_r + \frac{1}{r}\psi_{\theta,\theta} + \frac{1}{r}w_{,r} + \frac{1}{r^2}w_{,\theta\theta} + \nu^k(\psi_{r,r} + w_{,rr})\right)(z_{k+1}^5 - z_k^5)\right)$$

$$M_{r\theta} = \frac{1}{2}\sum_{k=1}^{N} \frac{E^k}{1+(\nu^k)}\left(\left(\frac{1}{r}u_{r,\theta} - \frac{1}{r}u_\theta + u_{\theta,r}\right)(z_{k+1}^2 - z_k^2) + \frac{1}{3}\left(\frac{1}{r}\psi_{r,\theta} - \frac{1}{r}\psi_\theta + \psi_{\theta,r}\right)(z_{k+1}^3 - z_k^3)\right.$$

$$\left. -\frac{4}{15h^2}\left(\frac{1}{r}\psi_{r,\theta} - \frac{1}{r}\psi_\theta + \psi_{\theta,r} - \frac{2}{r^2}w_{,\theta} - \frac{2}{r}w_{,r\theta}\right)(z_{k+1}^5 - z_k^5)\right)$$

$$P_{rr} = \sum_{k=1}^{N} \frac{E^k}{1-(\nu^k)^2}\left(\frac{1}{4}\left(u_{r,r} + \nu^k\left(\frac{1}{r}u_r + \frac{1}{r}u_{\theta,\theta}\right)\right)(z_{k+1}^4 - z_k^4)\right.$$

$$\left. +\frac{1}{5}\left(\psi_{r,r} + \nu^k\left(\frac{1}{r}\psi_r + \frac{1}{r}\psi_{\theta,\theta}\right)\right)(z_{k+1}^5 - z_k^5)\right.$$

(14)

$$\left. -\frac{4}{21h^2}\left(\psi_{r,r} + w_{,rr} + \nu^k\left(\frac{1}{r}\psi_r + \frac{1}{r}\psi_{\theta,\theta} + \frac{1}{r}w_{,r} + \frac{1}{r^2}w_{,\theta\theta}\right)\right)(z_{k+1}^7 - z_k^7)\right)$$

$$P_{\theta\theta} = \sum_{k=1}^{N} \frac{E^k}{1-(\nu^k)^2} \left( \frac{1}{4}\left( \nu^k u_{,r} + \frac{1}{r}u_{,r} + \frac{1}{r}u_{\theta,\theta} \right)\left( z_{k+1}^4 - z_k^4 \right) + \frac{1}{5}\left( \nu^k \psi_{,r} + \frac{1}{r}\psi_r + \frac{1}{r}\psi_{\theta,\theta} \right)\left( z_{k+1}^5 - z_k^5 \right) \right.$$

$$\left. -\frac{4}{21h^2}\left( \nu^k(\psi_{,r} + w_{,rr}) + \frac{1}{r}\psi_r + \frac{1}{r}\psi_{\theta,\theta} + \frac{1}{r}w_{,r} + \frac{1}{r^2}w_{,\theta\theta} \right)\left( z_{k+1}^7 - z_k^7 \right) \right) \quad (15)$$

$$P_{r\theta} = \frac{1}{2}\sum_{k=1}^{N} \frac{E^k}{1+\nu^k} \left( \frac{1}{4}\left( \frac{1}{r}u_{,\theta} - \frac{1}{r}u_\theta + u_{\theta,r} \right)(z_{k+1}^4 - z_k^4) + \frac{1}{5}\left( \frac{1}{r}\psi_{r,\theta} - \frac{1}{r}\psi_\theta + \psi_{\theta,r} \right)(z_{k+1}^5 - z_k^5) \right.$$

$$\left. -\frac{4}{21h^2}\left( \frac{1}{r}\psi_{r,\theta} - \frac{1}{r}\psi_\theta + \psi_{\theta,r} - \frac{2}{r^2}w_{,\theta} + \frac{2}{r}w_{,r\theta} \right)(z_{k+1}^7 - z_k^7) \right)$$

and,

$$Q_r = \sum_{k=1}^{N} G_3^k \left( (\psi_r + w_{,r})(z_{k+1} - z_k) - \frac{4}{3h^2}(\psi_r + w_{,r})(z_{k+1}^3 - z_k^3) \right)$$

$$Q_\theta = \sum_{k=1}^{N} G_3^k \left( \left( \psi_\theta + \frac{1}{r}w_{,\theta} \right)(z_{k+1} - z_k) - \frac{4}{3h^2}\left( \psi_\theta + \frac{1}{r}w_{,\theta} \right)(z_{k+1}^3 - z_k^3) \right) \quad (16)$$

and,

$$R_r = \sum_{k=1}^{N} G_3^k \left( \frac{1}{3}(\psi_r + w_{,r})(z_{k+1}^3 - z_k^3) - \frac{4}{5h^2}(\psi_r + w_{,r})(z_{k+1}^5 - z_k^5) \right)$$

$$R_\theta = \sum_{k=1}^{N} G_3^k \left( \frac{1}{3}\left( \psi_\theta + \frac{1}{r}w_{,\theta} \right)(z_{k+1}^3 - z_k^3) - \frac{4}{5h^2}\left( \psi_\theta + \frac{1}{r}w_{,\theta} \right)(z_{k+1}^5 - z_k^5) \right) \quad (17)$$

substituting Eqs. (13)–(16) into Eqs. (8) yields,

$$\overline{\alpha_1}\left( -\frac{1}{r^2}u + \frac{1}{r}u_{,r} + u_{,rr} - \frac{1}{r^2}v_{,\theta} + \frac{1}{r}v_{,r\theta} \right) + \overline{\alpha_2}\left( \frac{1}{r^2}u_{,\theta\theta} + v_{,\theta} - v_{,r\theta} \right)$$

$$+ \overline{\alpha_3}\left( -\frac{1}{r^2}\psi_r - \frac{1}{r}\psi_{r,r} + \psi_{r,rr} \right.$$

$$+\frac{1}{r^2}\psi_{\theta,\theta}+\frac{1}{r}\psi_{\theta,r\theta}\Big)+\overline{\alpha_4}\Big(\psi_{\theta,\theta}+\frac{1}{r^2}\psi_{r,\theta\theta}+\psi_{\theta,r\theta}\Big)$$
$$+\overline{\alpha_5}\Big(\frac{2}{r^3}w_{,\theta\theta}-\frac{1}{r}w_{,rr}-w_{,rrr}-\frac{1}{r^2}w_{,r\theta\theta}+\frac{1}{r^2}w_{,r}\Big)$$

$$=\overline{I_1}\ddot{u}+\overline{I_{22}}\ddot{\psi}_r-\overline{I_{44}}\ddot{w}_{,r}$$

$$\overline{\alpha_1}\Big(\frac{1}{r}u_{,r\theta}+\frac{1}{r^2}u_{,\theta}+\frac{1}{r^2}v_{,\theta\theta}\Big)+\overline{\alpha_2}\Big(-u_{,r\theta}+u_{,\theta}-\frac{1}{r^2}v+\frac{1}{r}v_{,r}+v_{,rr}\Big)+\overline{\alpha_3}\Big(\frac{1}{4r^2}\psi_{r,\theta}+\frac{2}{4r}\psi_{r,r\theta}$$

(18)

$$+\frac{1}{r^2}\psi_{\theta,\theta\theta}\Big)+\overline{\alpha_4}\Big(\frac{1}{4r}\psi_{r,r\theta}+\frac{1}{4r^2}\psi_{r,\theta}-\frac{1}{r^2}\psi_\theta+\frac{1}{r}\psi_{\theta,r}+\psi_{\theta,rr}\Big)$$
$$+\overline{\alpha_5}\Big(-\frac{1}{r^3}w_{,\theta\theta\theta}-\frac{1}{r}w_{,\theta rr}-\frac{1}{r^2}w_{,r\theta}\Big)$$

$$=\overline{I_1}\ddot{v}+\overline{I_{22}}\ddot{\psi}_\theta-\frac{1}{r}\overline{I_{44}}\ddot{w}_{,\theta}$$

$$\overline{\overline{\alpha_3}}\Big(-\frac{1}{r^2}u+\frac{1}{r}u_{,r}+u_{,rr}-\frac{1}{r^2}v_{,\theta}+\frac{1}{r}v_{,r\theta}\Big)+\overline{\overline{\alpha_4}}\Big(\frac{1}{r^2}u_{,\theta\theta}+v_{,\theta}-v_{,r\theta}\Big)+\overline{\overline{\alpha_6}}\Big(\frac{1}{3r^2}\psi_r-\frac{1}{3r}\psi_{r,r}+\frac{1}{r}\psi_{\theta,r\theta}-\frac{1}{3}\psi_{r,rr}$$
$$-\frac{1}{r^2}\psi_{\theta,\theta}\Big)+\overline{\overline{\alpha_7}}\Big(-\frac{1}{6r^2}\psi_{r,\theta\theta}+\psi_{\theta,\theta}-\psi_{\theta,r\theta}\Big)+\overline{\overline{\alpha_8}}\Big(-w_{,r}+\psi_r\Big)+\overline{\overline{\alpha_9}}\Big(\frac{2}{r^3}w_{,\theta\theta}-\frac{1}{r}w_{,rr}-w_{,rrr}-\frac{1}{r^2}w_{,r\theta\theta}+\frac{1}{r^2}w_{,r}\Big)$$
$$=\overline{\overline{I_{22}}}\ddot{u}+\overline{\overline{I_{33}}}\ddot{\psi}_r-\overline{\overline{I_{55}}}\ddot{w}_{,r}$$
$$\overline{\alpha_3}\Big(\frac{1}{r^2}u_{,\theta}+\frac{1}{r}u_{,r\theta}+\frac{1}{r^2}v_{,\theta\theta}\Big)+\overline{\alpha_4}\Big(\frac{1}{r^2}u_{,\theta}-\frac{1}{r}u_{,r\theta}-\frac{1}{r^2}v+\frac{1}{r}v_{,r}\Big)+\overline{\alpha_6}\Big(\frac{1}{r}\psi_{r,r\theta}+$$
$$\frac{1}{r^2}\psi_{r,\theta}+\frac{1}{r^2}\psi_{\theta,\theta\theta}\Big)+$$

$$\overline{\alpha_7}\Big(-\frac{1}{r}\psi_{r,r\theta}+\frac{1}{r^2}\psi_{r,\theta}-\frac{1}{r^2}\psi_\theta+\frac{1}{r}\psi_{\theta,r}+\psi_{\theta,rr}\Big)+\overline{\alpha_8}\Big(-\psi_\theta-\frac{1}{r}w_{,\theta}\Big)$$
$$+\overline{\alpha_9}\Big(-\frac{1}{r^2}w_{,r\theta}-\frac{1}{r^3}w_{,\theta\theta\theta}-\frac{1}{r}w_{,rr\theta}\Big)$$

$$=\overline{I_{22}}\ddot{v}+\overline{I_{33}}\ddot{\psi}_\theta-\frac{1}{r}\overline{I_{55}}\ddot{w}_{,\theta}$$

$$\overline{\alpha_5}\left(\frac{1}{r^3}u+\frac{1}{r^3}u_{,\theta\theta}-\frac{1}{r^2}u_{,r}+\frac{2}{r}u_{,rr}+\frac{1}{r}v_{,rr\theta}+\frac{1}{r^3}v_{,\theta}-\frac{1}{r^2}v_{,r\theta}+\frac{1}{r^2}u_{,r\theta\theta}+\frac{1}{r^2}v_{,\theta\theta\theta}\right.$$
$$\left.+u_{,rrr}\right)+\overline{\alpha_8}\left(\psi_{r,r}+\frac{1}{r}\psi_r+\frac{1}{r}w_{,r}\right.$$

$$+\frac{1}{r^2}w_{,\theta\theta}+w_{,rr}+\frac{1}{r}\psi_{\theta,\theta}\right)$$
$$+\overline{\alpha_9}\left(\frac{1}{r^2}\psi_{r,r\theta\theta}+\frac{1}{r^2}\psi_{r,r}+\frac{1}{r^3}\psi_r+\frac{1}{r^3}\psi_{\theta,\theta}+\frac{1}{r^3}\psi_{\theta,\theta\theta\theta}+\frac{1}{r}\psi_{r,rr}\right.$$
$$\left.+\frac{1}{r}\psi_{\theta,rr\theta}+\psi_{r,rrr}\right)$$

$$+\frac{1}{r^3}\psi_{r,\theta\theta}\right)+\overline{\alpha_{10}}\left(\frac{2}{r^3}w_{,r\theta\theta}-\frac{1}{r^4}w_{,\theta\theta\theta\theta}-\frac{2}{r^2}w_{,rr\theta\theta}-\frac{2}{r}w_{,rrr}-\frac{1}{r^3}w_{,r}-w_{,rrrr}\right.$$
$$\left.-\frac{4}{r^4}w_{,\theta\theta}+\frac{1}{r^2}w_{,rr}\right)$$

$$=\overline{I_{11}}\ddot{w}+\overline{I_{44}}(\ddot{u}_{,r}+\frac{1}{r}\ddot{v}_{,\theta})+\overline{I_{55}}(\ddot{\psi}_{r,r}+\frac{1}{r}\ddot{\psi}_{\theta,\theta}+\frac{1}{r}\ddot{\psi}_r)-\overline{I_{77}}\nabla^2\ddot{w}$$

### 4. Solution

where the coefficients $\alpha_i(i=1,2,\ldots,10)$ and $I_i(i=1,2,3,4,5,7)$ are listed in Appendix A.

$$\overline{\alpha_3}=\overline{\alpha_4}=\overline{\alpha_5}=\overline{I_{22}}=\overline{I_{44}}=0$$
(19)

Hence, the governing equations (18) are simplified as

$$\overline{\alpha_1}\left(-\frac{1}{r^2}u+\frac{1}{r}u_{,r}+u_{,rr}-\frac{1}{r^2}v_{,\theta}+\frac{1}{r}v_{,r\theta}\right)+\overline{\alpha_2}\left(\frac{1}{r^2}u_{,\theta\theta}+v_{,\theta}-v_{,r\theta}\right)=\overline{I_1}\ddot{u}$$

$$\overline{\alpha_1}\left(\frac{1}{r}u_{,r\theta}+\frac{1}{r^2}u_{,\theta}+\frac{1}{r^2}v_{,\theta\theta}\right)+\overline{\alpha_2}\left(-u_{,r\theta}+u_{,\theta}-\frac{1}{r^2}v+\frac{1}{r}v_{,r}+v_{,rr}\right)=\overline{I_1}\ddot{v}$$
(20)

$$\overline{\alpha_6}\left(\frac{1}{3r^2}\psi_r - \frac{1}{3r}\psi_{r,r} + \frac{1}{r}\psi_{\theta,r\theta} - \frac{1}{3}\psi_{r,rr} - \frac{1}{r^2}\psi_{\theta,\theta}\right) + \overline{\alpha_7}\left(-\frac{1}{6r^2}\psi_{r,\theta\theta} + \psi_{\theta,\theta} - \psi_{\theta,r\theta}\right) + \overline{\alpha_8}\left(-w_{,r} + \psi_r\right)$$

$$+\overline{\alpha_9}\left(\frac{2}{r^3}w_{,\theta\theta} - \frac{1}{r}w_{,rr} - w_{,rrr} - \frac{1}{r^2}w_{,r\theta\theta} + \frac{1}{r^2}w_{,r}\right) = \overline{I_{33}}\ddot{\psi}_r - \overline{I_{55}}\ddot{w}_{,r}$$

$$\overline{\alpha_6}\left(\frac{1}{r}\psi_{r,r\theta} + \frac{1}{r^2}\psi_{r,\theta} + \frac{1}{r^2}\psi_{\theta,\theta\theta}\right) + \overline{\alpha_7}\left(-\frac{1}{r}\psi_{r,r\theta} + \frac{1}{r^2}\psi_{r,\theta} - \frac{1}{r^2}\psi_\theta + \frac{1}{r}\psi_{\theta,r} + \psi_{\theta,rr}\right) +$$

$$\overline{\alpha_8}\left(-\psi_\theta - \frac{1}{r}w_{,\theta}\right)$$

$$+\overline{\alpha_9}\left(-\frac{1}{r^2}w_{,r\theta} - \frac{1}{r^3}w_{,\theta\theta\theta} - \frac{1}{r}w_{,rr\theta}\right) = \overline{I_{33}}\ddot{\psi}_\theta - \frac{1}{r}\overline{I_{55}}\ddot{w}_{,\theta}$$

$$\overline{\alpha_8}\left(\psi_{r,r} + \frac{1}{r}\psi_r + \frac{1}{r}w_{,r} + \frac{1}{r^2}w_{,\theta\theta} + w_{,rr} + \frac{1}{r}\psi_{\theta,\theta}\right)$$

$$+\overline{\alpha_9}\left(\frac{1}{r^2}\psi_{r,r\theta\theta} + \frac{1}{r^2}\psi_{r,r} + \frac{1}{r^3}\psi_r + \frac{1}{r^3}\psi_{\theta,\theta} + \frac{1}{r^3}\psi_{\theta,\theta\theta\theta}\right.$$

$$+\frac{1}{r}\psi_{r,rr} + \frac{1}{r}\psi_{\theta,rr\theta} + \psi_{r,rrr} + \frac{1}{r^3}\psi_{r,\theta\theta}\right)$$

$$+\overline{\alpha_{10}}\left(\frac{2}{r^3}w_{,r\theta\theta} - \frac{1}{r^4}w_{,\theta\theta\theta\theta} - \frac{2}{r^2}w_{,rr\theta\theta} - \frac{2}{r}w_{,rrr} - \frac{1}{r^3}w_{,r} - w_{,rrrr}\right)$$

$$-\frac{4}{r^4}w_{,\theta\theta} + \frac{1}{r^2}w_{,rr}\right) = \overline{I_{11}}\ddot{w} + \overline{I_{55}}(\ddot{\psi}_{r,r} + \frac{1}{r}\ddot{\psi}_{\theta,\theta} + \frac{1}{r}\ddot{\psi}_r) - \overline{I_{77}}\nabla^2\ddot{w}$$

the displacement and rotation components assume the following forms,

$$u = \tilde{u}\exp(j\omega t)$$
$$v = \tilde{v}\exp(j\omega t)$$
$$w = \tilde{w}\exp(j\omega t)$$
$$\psi_r = \tilde{\psi}_r\exp(j\omega t)$$
$$\psi_\theta = \tilde{\psi}_\theta\exp(j\omega t) \qquad j = \sqrt{-1}$$

(21)

where $\omega$ denotes the frequency of vibration. Substituting Eqs. (21) into Eqs. (20) yields,

$$\overline{\alpha_1}\left(-\frac{1}{r^2}\tilde{u} + \frac{1}{r}\tilde{u}_{,r} + \tilde{u}_{,rr} - \frac{1}{r^2}\tilde{v}_{,\theta} + \frac{1}{r}\tilde{v}_{,r\theta}\right) + \overline{\alpha_2}\left(\frac{1}{r^2}\tilde{u}_{,\theta\theta} + \tilde{v}_{,\theta} - \tilde{v}_{,r\theta}\right) + \overline{I_{11}}\omega^2\tilde{u}$$

(22)

$$\overline{\alpha_1}\left(\frac{1}{r}\tilde{u}_{,r\theta}+\frac{1}{r^2}\tilde{u}_{,\theta}+\frac{1}{r^2}\tilde{v}_{,\theta\theta}\right)+\overline{\alpha_2}\left(-\tilde{u}_{,r\theta}+\tilde{u}_{,\theta}-\frac{1}{r^2}\tilde{v}+\frac{1}{r}\tilde{v}_{,r}+\tilde{v}_{,rr}\right)+\overline{I_1}\omega^2\tilde{v}=0$$

$$\overline{\overline{\alpha_6}}\left(\frac{1}{3r^2}\tilde{\psi}_r-\frac{1}{3r}\tilde{\psi}_{r,r}+\frac{1}{r}\tilde{\psi}_{\theta,r\theta}-\frac{1}{3}\tilde{\psi}_{r,rr}-\frac{1}{r^2}\tilde{\psi}_{\theta,\theta}\right)+\overline{\overline{\alpha_7}}\left(-\frac{1}{6r^2}\tilde{\psi}_{r,\theta\theta}+\tilde{\psi}_{\theta,\theta}-\tilde{\psi}_{\theta,r\theta}\right)+\overline{\overline{\alpha_8}}\left(-\tilde{w}_{,r}+\tilde{\psi}_r\right)$$

$$+\overline{\overline{\alpha_9}}\left(\frac{2}{r^3}\tilde{w}_{,\theta\theta}-\frac{1}{r}\tilde{w}_{,rr}-\tilde{w}_{,rrr}-\frac{1}{r^2}\tilde{w}_{,r\theta\theta}+\frac{1}{r^2}\tilde{w}_{,r}\right)+\overline{\overline{I_{33}}}\omega^2\tilde{\psi}_r-\overline{\overline{I_{55}}}\omega^2\tilde{w}_{,r}=0$$

$$\overline{\alpha_6}\left(\frac{1}{r}\tilde{\psi}_{r,r\theta}+\frac{1}{r^2}\tilde{\psi}_{r,\theta}+\frac{1}{r^2}\tilde{\psi}_{\theta,\theta\theta}\right)+\overline{\alpha_7}\left(-\frac{1}{r}\tilde{\psi}_{r,r\theta}+\frac{1}{r^2}\tilde{\psi}_{r,\theta}-\frac{1}{r^2}\tilde{\psi}_\theta+\frac{1}{r}\tilde{\psi}_{\theta,r}+\tilde{\psi}_{\theta,rr}\right)+$$
$$\overline{\alpha_8}\left(-\tilde{\psi}_\theta-\frac{1}{r}\tilde{w}_{,\theta}\right)$$

$$+\overline{\alpha_9}\left(-\frac{1}{r^2}\tilde{w}_{,r\theta}-\frac{1}{r^3}\tilde{w}_{,\theta\theta\theta}-\frac{1}{r}\tilde{w}_{,rr\theta}\right)+\overline{I_{33}}\omega^2\tilde{\psi}_\theta-\frac{1}{r}\overline{I_{55}}\omega^2\tilde{w}_{,\theta}=0$$

$$\overline{\alpha_8}\left(\tilde{\psi}_{r,r}+\frac{1}{r}\tilde{\psi}_r+\frac{1}{r}\tilde{w}_{,r}+\frac{1}{r^2}\tilde{w}_{,\theta\theta}+\tilde{w}_{,rr}+\frac{1}{r}\tilde{\psi}_{\theta,\theta}\right) \qquad (23)$$

$$+\overline{\alpha_9}\left(\frac{1}{r^2}\tilde{\psi}_{r,r\theta\theta}+\frac{1}{r^2}\tilde{\psi}_{r,r}+\frac{1}{r^3}\tilde{\psi}_r+\frac{1}{r^3}\tilde{\psi}_{\theta,\theta}+\frac{1}{r^3}\tilde{\psi}_{\theta,\theta\theta\theta}\right.$$

$$\left.+\frac{1}{r}\tilde{\psi}_{r,rr}+\frac{1}{r}\tilde{\psi}_{\theta,rr\theta}+\tilde{\psi}_{r,rrr}+\frac{1}{r^3}\tilde{\psi}_{r,\theta\theta}\right)$$
$$+\overline{\alpha_{10}}\left(\frac{2}{r^3}\tilde{w}_{,r\theta\theta}-\frac{1}{r^4}\tilde{w}_{,\theta\theta\theta\theta}-\frac{2}{r^2}\tilde{w}_{,rr\theta\theta}-\frac{2}{r}\tilde{w}_{,rrr}-\frac{1}{r^3}\tilde{w}_{,r}-\tilde{w}_{,rrrr}\right.$$

$$\left.-\frac{4}{r^4}\tilde{w}_{,\theta\theta}+\frac{1}{r^2}\tilde{w}_{,rr}\right)+\overline{I_{11}}\omega^2\tilde{w}+\overline{I_{55}}\omega^2(\tilde{\psi}_{r,r}+\frac{1}{r}\tilde{\psi}_{\theta,\theta}+\frac{1}{r}\tilde{\psi}_r)-\overline{I_{77}}\omega^2\nabla^2\tilde{w}=0$$

$$\tilde{\eta}=\frac{1}{r}\tilde{u}_{,\theta}-\tilde{v}_{,r}-\frac{1}{r}\tilde{v}$$

$$\tilde{\mu}=\tilde{u}_{,r}+\frac{1}{r}\tilde{u}+\frac{1}{r}\tilde{v}_{,\theta}$$

$$\tilde{\varphi}=\frac{1}{r}\tilde{\psi}_{r,\theta}-\tilde{\psi}_{\theta,r}-\frac{1}{r}\tilde{\psi}_\theta$$

$$\overline{\alpha_1}\left(\tilde{u}_{,r}+\frac{1}{r}\tilde{u}+\frac{1}{r}\tilde{v}_{,\theta}\right)_{,r}+\overline{\alpha_2}\left(\frac{1}{r^2}\tilde{u}_{,\theta\theta}+\tilde{v}_{,\theta}-\tilde{v}_{,r\theta}\right)+\overline{I_{11}}\omega^2\tilde{u}$$

$$\overline{\alpha_1}\left(\frac{1}{r}\left(\tilde{u}_{,r}+\frac{1}{r}\tilde{u}+\frac{1}{r}\tilde{v}_{,\theta}\right)_{,r}\right)+\overline{\alpha_2}\left(-\tilde{u}_{,r\theta}+\tilde{u}_{,\theta}-\frac{1}{r^2}\tilde{v}+\frac{1}{r}\tilde{v}_{,r}+\tilde{v}_{,rr}\right)+\overline{I_1}\omega^2\tilde{v}=0 \qquad (24)$$

$$\overline{\overline{\alpha_6}}\left(\frac{1}{3r^2}\tilde{\psi}_{,r}-\frac{1}{3r}\tilde{\psi}_{,rr}+\frac{1}{r}\tilde{\psi}_{,\theta,r\theta}-\frac{1}{3}\tilde{\psi}_{,rrr}-\frac{1}{r^2}\tilde{\psi}_{,\theta\theta}\right)+\overline{\overline{\alpha_7}}\left(-\frac{1}{6r^2}\tilde{\psi}_{,r\theta}+\tilde{\psi}_{,\theta\theta}-\tilde{\psi}_{,\theta,r\theta}\right)+\overline{\overline{\alpha_8}}\left(-\tilde{w}_{,r}+\tilde{\psi}_{,r}\right)$$

$$+\overline{\overline{\alpha_9}}\left(\frac{2}{r^3}\tilde{w}_{,\theta\theta}-\frac{1}{r}\tilde{w}_{,rr}-\tilde{w}_{,rrr}-\frac{1}{r^2}\tilde{w}_{,r\theta\theta}+\frac{1}{r^2}\tilde{w}_{,r}\right)+\overline{I_{33}}\omega^2\tilde{\psi}_{,r}-\overline{I_{55}}\omega^2\tilde{w}_{,r}=0$$

$$\overline{\alpha_6}\left(\frac{1}{r}\tilde{\psi}_{,r\theta}+\frac{1}{r^2}\tilde{\psi}_{,r\theta}+\frac{1}{r^2}\tilde{\psi}_{,\theta\theta}\right)+\overline{\alpha_7}\left(-\frac{1}{r}\tilde{\psi}_{,r\theta}+\frac{1}{r^2}\tilde{\psi}_{,r\theta}-\frac{1}{r^2}\tilde{\psi}_{,\theta}+\frac{1}{r}\tilde{\psi}_{,\theta,r}+\tilde{\psi}_{,\theta,rr}\right)+$$
$$\overline{\alpha_8}\left(-\tilde{\psi}_{,\theta}-\frac{1}{r}\tilde{w}_{,\theta}\right)$$

$$+\overline{\alpha_9}\left(-\frac{1}{r^2}\tilde{w}_{,r\theta}-\frac{1}{r^3}\tilde{w}_{,\theta\theta\theta}-\frac{1}{r}\tilde{w}_{,rr\theta}\right)+\overline{I_{33}}\omega^2\tilde{\psi}_{,\theta}-\frac{1}{r}\overline{I_{55}}\omega^2\tilde{w}_{,\theta}=0$$

$$\overline{\alpha_8}\left(\tilde{\psi}_{,r,r}+\frac{1}{r}\tilde{\psi}_{,r}+\frac{1}{r}\tilde{w}_{,r}+\frac{1}{r^2}\tilde{w}_{,\theta\theta}+\tilde{w}_{,rr}+\frac{1}{r}\tilde{\psi}_{,\theta,\theta}\right)$$
$$+\overline{\alpha_9}\left(\frac{1}{r^2}\tilde{\psi}_{,r,r\theta\theta}+\frac{1}{r^2}\tilde{\psi}_{,r,r}+\frac{1}{r^3}\tilde{\psi}_{,r}+\frac{1}{r^3}\tilde{\psi}_{,\theta,\theta}+\frac{1}{r^3}\tilde{\psi}_{,\theta,\theta\theta}\right)$$

$$+\frac{1}{r}\tilde{\psi}_{,r,rr}+\frac{1}{r}\tilde{\psi}_{,\theta,rr\theta}+\tilde{\psi}_{,r,rrr}+\frac{1}{r^3}\tilde{\psi}_{,r,\theta\theta}\right)$$
$$+\overline{\alpha_{10}}\left(\frac{2}{r^3}\tilde{w}_{,r\theta}-\frac{1}{r^4}\tilde{w}_{,\theta\theta\theta\theta}-\frac{2}{r^2}\tilde{w}_{,rr\theta\theta}-\frac{2}{r}\tilde{w}_{,rrr}-\frac{1}{r^3}\tilde{w}_{,r}-\tilde{w}_{,rrrr}\right.$$

$$\left.-\frac{4}{r^4}\tilde{w}_{,\theta\theta}+\frac{1}{r^2}\tilde{w}_{,rr}\right)+\overline{I_{11}}\omega^2\tilde{w}+\overline{I_{55}}\omega^2\left(\tilde{\psi}_{,r,r}+\frac{1}{r}\tilde{\psi}_{,\theta,\theta}+\frac{1}{r}\tilde{\psi}_{,r}\right)-\overline{I_{77}}\omega^2\nabla^2\tilde{w}=0$$

five coupled Eqs. (23b) can be converted into four uncoupled as follows,

$$\overline{\ell_1}\nabla^2\tilde{\eta}+\overline{\ell_2}\tilde{\eta}=0 \qquad (25a)$$

$$\overline{\ell_3}\nabla^2\tilde{\mu}+\overline{\ell_4}\tilde{\mu}=0 \qquad (25b)$$

$$\overline{\ell_5}\nabla^2\tilde{\varphi}-\overline{\ell_6}\tilde{\varphi}=0 \qquad (25c)$$

$$\overline{\ell_7}\nabla^6\tilde{w}+\overline{\ell_8}\nabla^4\tilde{w}-\overline{\ell_9}\nabla^2\tilde{w}-\overline{\ell_{10}}\tilde{w}=0 \qquad (25d)$$

where the coefficients $\overline{\ell_1}$ through $\overline{\ell_{10}}$ are given in Appendix B. where $\nabla^2$ is the two-dimensional Laplace operator in polar coordinates.

$$\nabla^2 = \frac{\partial^2}{\partial r^2} + \frac{\partial}{r\partial r} + \frac{1}{r^2}\frac{\partial^2}{\partial \theta^2}$$

Radial and circumferential displacements can be expressed as,

$$\tilde{u} = -\frac{1}{\overline{I_{11}}\omega^2}\left(\overline{\alpha_1}\tilde{\mu}_{,r} + \frac{1}{r}\overline{\alpha_2}\tilde{\eta}_{,\theta}\right)$$

$$\tilde{v} = -\frac{1}{\overline{I_{11}}\omega^2}\left(\frac{1}{r}\overline{\alpha_1}\tilde{\mu}_{,\theta} - \overline{\alpha_2}\tilde{\eta}_{,r}\right) \tag{26}$$

It is easy to show that the rotation functions $\psi_r, \psi_\theta$ can be written in terms of transverse deflection $w$ and $\varphi$

$$\tilde{\psi}_r = \frac{1}{\overline{\alpha_8} - \overline{I_{33}}\omega^2}\left(\overline{\Lambda_1}\nabla^4\tilde{w}_{,r} - \overline{\Lambda_2}\nabla^2\tilde{w}_{,r} - \overline{\Lambda_3}\tilde{w}_{,r} + \frac{1}{r}\overline{\Lambda_4}\tilde{\varphi}_{,\theta}\right)$$

$$\tilde{\psi}_\theta = \frac{1}{\overline{\alpha_8} - \overline{I_{33}}\omega^2}\left(\frac{1}{r}\overline{\Lambda_1}\nabla^4\tilde{w}_{,\theta} - \frac{1}{r}\overline{\Lambda_2}\nabla^2\tilde{w}_{,\theta} - \frac{1}{r}\overline{\Lambda_3}\tilde{w}_{,\theta} - \overline{\Lambda_4}\tilde{\varphi}_{,r}\right) \tag{27}$$

where,

$$\overline{\Lambda_1} = \frac{\overline{\alpha_6}\left(\overline{\alpha_6}\overline{\alpha_{10}} - \overline{\alpha_9}^2\right)}{\overline{\alpha_8}(\overline{\alpha_6} + \overline{\alpha_9}) - (\overline{\alpha_9}\overline{I_{33}} - \overline{\alpha_6}\overline{I_{55}})\omega^2}$$

$$\overline{\Lambda_2} = \frac{\overline{\alpha_8}(\overline{\alpha_6}+\overline{\alpha_9})^2 - \left(\overline{\alpha_9}^2\overline{I_{33}} - 2\overline{\alpha_9}\overline{\alpha_6}\overline{I_{55}} + \overline{\alpha_6}^2\overline{I_{77}}\right)\omega^2}{\overline{\alpha_8}(\overline{\alpha_6}+\overline{\alpha_9}) - (\overline{\alpha_9}\overline{I_{33}} - \overline{\alpha_6}\overline{I_{55}})\omega^2}$$

$$\overline{\Lambda_3}$$
$$= \frac{\overline{\alpha_8}^2(\overline{\alpha_6} + \overline{\alpha_9}) + \left(\overline{\alpha_6}^2\overline{I_{11}} - \overline{\alpha_8}\overline{\alpha_9}\overline{I_{33}} + \overline{\alpha_8}(\overline{\alpha_9} + 2\overline{\alpha_6})\overline{I_{55}}\right)\omega^2 - \overline{I_{55}}^2\left(\overline{\alpha_9}\overline{I_{33}} - \overline{\alpha_6}\overline{I_{55}}\right)\omega^4}{\overline{\alpha_8}(\overline{\alpha_6} + \overline{\alpha_9}) - (\overline{\alpha_9}\overline{I_{33}} - \overline{\alpha_6}\overline{I_{55}})\omega^2}$$

$$\overline{\Lambda_4} = \overline{\alpha_7}$$

(28)

*4.1. Solution for u,v*

To obtain u, v the values of the parameters of the equation(26) must $\eta, \mu$ first be determined by the equation:

$$\overline{\ell_1}\nabla^2\tilde{\eta} + \overline{\ell_2}\tilde{\eta} = 0 \tag{29}$$

$$\overline{\ell_3}\nabla^2\tilde{\mu} + \overline{\ell_4}\tilde{\mu} = 0$$

$$\nabla^2 = \frac{\partial^2}{\partial r^2} + \frac{\partial}{r\partial r} + \frac{1}{r^2}\frac{\partial^2}{\partial \theta^2} \tag{30}$$

The new constants are defined as follows,

$$y_1 = \xi_1^{\,2} = \frac{\overline{\ell_2}}{\overline{\ell_1}}$$

$$y_2 = \xi_2^{\,2} = \frac{\overline{\ell_4}}{\overline{\ell_3}} \tag{31}$$

Equations can be rewritten as follows:

$$\nabla^2\tilde{\eta} + \xi_1^{\,2}\tilde{\eta} = 0 \tag{32}$$

$$\nabla^2\tilde{\mu} + \xi_2^{\,2}\tilde{\mu} = 0$$

the potential functions $\tilde{\eta}, \tilde{\mu}$ can be represented by

$$\tilde{\eta} = \eta(r)\cos(p\theta) \tag{33}$$

$$\tilde{\mu} = \mu(r)\sin(p\theta) \tag{34}$$

$$\eta''(r) + \frac{1}{r}\eta'(r) + (\xi_1^2 - \frac{1}{r^2}p^2)\eta(r) = 0 \tag{35}$$

$$\mu''(r) + \frac{1}{r}\mu'(r) + (\xi_2^2 - \frac{1}{r^2}p^2)\mu(r) = 0 \tag{36}$$

$$\eta(r) = \sum_{i=1}^{3}(A_{i1}\eta_{i1}(p,\xi_i r) + A_{i2}\eta_{i2}(p,\xi_i r)) \tag{37}$$

$$\eta_{i1}(p,\xi_1 r) = \begin{cases} J_p(\xi_i r), \\ I_p(\xi_i r), \end{cases} \quad y_i < 0 \tag{38}$$

$y_i > 0$

and ,

$$\eta_{i2}(p,\xi_i r)=\begin{cases}Y_p(\xi_i r),\\ K_p(\xi_i r),\end{cases}\quad y_i<0 \tag{39}$$

$y_i>0$

$$\mu(r)=\sum_{i=1}^{3}(B_{i1}\mu(p,\xi_i r)+B_{i2}\mu(p,\xi_i r)) \tag{40}$$

$$\mu(p,\xi_2 r)=\begin{cases}J_p(\xi_2 r),\\ I_p(\xi_2 r),\end{cases}\quad y_i<0 \tag{41}$$

$y_i>0$

$$\mu(p,\xi_i r)=\begin{cases}Y_p(\xi_i r),\\ K_p(\xi_i r),\end{cases}\quad y_i<0 \tag{42}$$

$y_i>0$

*4.2. Solution for transverse displacement*

After some mathematical manipulation, the equation(25d) of motions can be rewritten as follows:

$$p_1\overline{\nabla\nabla\nabla}w+p_2\overline{\nabla\nabla}w+p_3\overline{\nabla}w+p_4=0 \tag{43}$$

$$\overline{\nabla}=\nabla^2=\frac{\partial^2}{\partial r^2}+\frac{\partial}{r\partial r}+\frac{1}{r^2}\frac{\partial^2}{\partial\theta^2} \tag{44}$$

in which the coefficients $p_i\ (i=1,2,3,4)$ are determined by

$$\begin{aligned}p_1&=\overline{\ell_7}\\ p_2&=\overline{\ell_8}\\ p_3&=-\overline{\ell_9}\\ p_4&=-\overline{\ell_{10}}\end{aligned} \tag{45}$$

where the coefficients $\overline{\ell_7}$ through $\overline{\ell_{10}}$ are given in Appendix B.

functions $w$ can be represented as form:

$$w(r,\theta) = \tilde{w}(r)\cos(p\theta) \tag{46}$$

where p is the circumferential wave number corresponding mode shape. The governing differential equations in Eqs. (1)– (3) may be transformed as,

$$(\bar{\nabla} - X_1)(\bar{\nabla} - X_2)(\bar{\nabla} - X_3)\tilde{w} = 0 \tag{47}$$

where $x_1$, $x_2$ and $x_3$ are the three roots of the following equation,

$$p_1 X^3 + p_2 X^2 + p_3 X + p_4 = 0 \tag{48}$$

The general solution to Eq. (30) may be expressed as,

$$\tilde{w} = \tilde{w}_1 + \tilde{w}_2 + \tilde{w}_3 \tag{49}$$

in which $\tilde{w}_i$ ($i = 1,2,3$) are obtained by three different kinds of Bessel's equations as follows,

$$(\bar{\nabla} - X_1)\tilde{w} = 0 \tag{50}$$

$$(\bar{\nabla} - X_2)\tilde{w} = 0 \tag{51}$$

$$(\bar{\nabla} - X_3)\tilde{w} = 0 \tag{52}$$

The second-order term of Eq. (31) can easily be eliminated by using the following Transformation,

$$X = Z - \frac{p_2}{3p_1} \tag{53}$$

Thus, Eq. (31) reduces to

$$Z^3 + sZ + t = 0 \tag{54}$$

Where,

$$s = \frac{p_3}{p_1} - \frac{p_2^2}{3p_1^2}, \qquad t = \frac{p_4}{p_1} - \frac{p_2 p_3}{3p_1^2} + \frac{2p_2^3}{27p_1^3} \tag{55}$$

It is well known that the discriminant of a third-order equation can be expressed as,

$$\delta = \left(\frac{t}{2}\right)^2 + \left(\frac{s}{3}\right)^3 \tag{56}$$

The parameter $\delta$ practically takes negative values $\delta < 0$. Therefore, based on Cardano's formula [23], three distinct real roots of Eq. (31) are given by

$$X_1 = 2\Gamma \cos\frac{\kappa}{3} - \frac{p_2}{3p_1} \tag{57}$$

$$X_2 = 2\Gamma \cos\frac{\kappa + 2\pi}{3} - \frac{p_2}{3p_1} \tag{58}$$

$$X_3 = 2\Gamma \cos\frac{\kappa + 4\pi}{3} - \frac{p_2}{3p_1} \tag{59}$$

where

$$\Gamma = \frac{1}{3}\sqrt{\frac{p_2^2 - 3p_1 p_3}{p_1^2}}, \qquad \kappa = \arccos\left[-\frac{t}{2\sqrt{\left(\frac{-s}{3}\right)^3}}\right] \tag{60}$$

The solution of Eq. (27) can be given by,

$$w(r,\theta) = \sum_{i=1}^{3}\left(c_{i1} w_{i1}(p, \lambda_i r) + c_{i2} w_{i2}(p, \lambda_i r)\right)\cos(p\theta) \tag{61}$$

Where,

$$\lambda_i = \sqrt{|X_i|} \tag{62}$$

$$w_{i1}(p, \lambda_i r) = \begin{cases} J_p(\lambda_i r), & x_i < 0 \\ I_P(\lambda_i r), \end{cases} \tag{63}$$

$x_i > 0 \quad i=1,2,3$

$$w_{i2}(p, \lambda_i r) = \begin{cases} Y_p(\lambda_i r), & x_i < 0 \\ K_P(\lambda_i r), \end{cases} \tag{64}$$

$x_i > 0 \quad i=1,2,3$

and $c_i (i = 1,2,3,4,5,6)$ are unknown coefficients; $J_p$ and $Y_p$ are the Bessel functions of the first and the second kind, respectively, whereas $I_p$ and $K_p$ are the modified Bessel functions of the first and the second kind, respectively.

the potential functions $\varphi$ can be represented by,

$$\varphi = \tilde{\varphi} \sin(p\theta) \tag{65}$$

$$\tilde{\varphi}(r) = \sum_{i=4}^{4}(c_{i1}\varphi_{i1}(p,\lambda_i r) + c_{i2}\varphi_{i2}(p,\lambda_i r)) \tag{66}$$

$$\varphi_{i1}(p,\lambda_i r) = \begin{cases} J_p(\lambda_i r), \\ I_P(\lambda_i r), \end{cases} \quad x_i < 0 \tag{67}$$

$x_i > 0 \quad i=4$

$$\varphi_{i2}(p,\lambda_i r) = \begin{cases} Y_p(\lambda_i r), \\ K_P(\lambda_i r), \end{cases} \quad x_i < 0 \tag{68}$$

$x_i > 0 \quad i=4$

The transverse displacement $w$ along with the slope rotations $\psi_r, \psi_\theta$, were exactly determined in terms of the frequency parameter $\beta$. Edges of the circular plate may take any classical boundary conditions, including free, soft simply supported, hard simply supported and clamped. The rotation functions can be obtained following forms as

$$\tilde{\psi}_r = \frac{1}{\alpha_8 - I_{33}\omega^2}\left(\overline{\Lambda}_1 \nabla^4 \tilde{w}_{,r} - \overline{\Lambda}_2 \nabla^2 \tilde{w}_{,r} - \overline{\Lambda}_3 \tilde{w}_{,r} + \frac{1}{r}\overline{\Lambda}_4 \tilde{\varphi}_{,\theta}\right)$$

$$\tilde{\psi}_\theta = \frac{1}{\alpha_8 - I_{33}\omega^2}\left(\frac{1}{r}\overline{\Lambda}_1 \nabla^4 \tilde{w}_{,\theta} - \frac{1}{r}\overline{\Lambda}_2 \nabla^2 \tilde{w}_{,\theta} - \frac{1}{r}\overline{\Lambda}_3 \tilde{w}_{,\theta} - \overline{\Lambda}_4 \tilde{\varphi}_{,r}\right) \tag{69}$$

## 5. Results and discussion

For numerical calculations, single-layer and three-layer transversely isotropic plate with the following non-dimensional properties are considered

*Single-layer plates:*

$$\bar{G} = 2.5, \quad \nu = 0.3 \tag{70}$$

*Three-layer plates:*

$$\frac{G^{(3)}}{G^{(1)}} = 1, \quad \frac{G^{(2)}}{G^{(1)}} = \frac{1}{2} \tag{71}$$

$$\frac{\rho^{(3)}}{\rho^{(1)}} = 1, \quad \frac{\rho^{(2)}}{\rho^{(1)}} = \frac{2}{3}$$

$$\nu^{(1)} = 0.35, \qquad \nu^{(2)} = 0.25, \quad \nu^{(3)} = 0.35$$

$$\bar{G}^{(1)} = 2, \qquad \bar{G}^{(2)} = 1.25, \quad \bar{G}^{(3)} = 2$$

where $\bar{\omega}_m|out$ and $\bar{\omega}_m|in$ are the non-dimensional out-of-plane and in-plane natural frequencies, are defined as follows,

$$\bar{\omega}_m\bigg|out = \left(\frac{a^2}{h}\right)\sqrt{\left(\frac{\rho}{G}\right)}\omega$$

$$\bar{\omega}_m|in = a\sqrt{\left(\frac{\rho}{G}\right)}\omega$$

$$\bar{h} = \frac{h}{a}$$

$$\bar{G} = \frac{G}{G_3}$$

Also, $h$ is thickness to radius, $G$ is shear modulus ratio. It is noticeable that $G$ and $G_3$ are the in-plane and out-of-plane shear modulus of the transversely isotropic material, respectively. Exact solutions for vibration analysis of thick circular- annular plates with different combinations of free, soft simply supported, hard simply supported and clamped outer and inner boundaries, are computed. The vibration frequency is expressed in terms of a non-dimensional frequency parameter. and circumferential distribution of the mode shapes corresponding to each boundary condition are depicted in Figs. 2, 3, 4, 5, 6, 7, 8 9,10 and 11. $\overline{\omega m}|_{out} = \left(\frac{a^2}{h}\right)\sqrt{\left(\frac{\rho}{G_z}\right)}\omega$, is shown in Tables 1 to 4 for thick circular plate with transversely isotropic material for various thickness–radius ratios (h/a) 1/50,1/10/,1/5 and 1/3 also The non-dimensional frequency parameter $\overline{\omega}_m|_{in} = a\sqrt{\left(\frac{\rho}{G_z}\right)}\omega$ is shown for dented plate in table 5,6 . and The non-dimensional frequency parameter $\bar{\omega}m|_{out} = \left(\frac{a^2}{h}\right)\sqrt{\left(\frac{\rho}{G_z}\right)}\omega, \bar{\omega}_m|_{in} = a\sqrt{\left(\frac{\rho}{G_z}\right)}\omega$ is shown in Tables 7 to 9 and Tables 11 to 13 for circular plate with laminate transversely isotropic material for various thickness–radius ratios (h/a) 0.3 and inner-outer radius ratios (b/a)=0.4.

The results computed by exact method are compared with the FEM method in Table 1-14. It can be found that the present method can obtain very high accurate frequencies of the thick circular plates.

for the non-dimensional frequency parameter out-of plane and in-plane $\bar{\omega}m|_{out} = \left(\frac{a^2}{h}\right)\sqrt{\left(\frac{\rho}{G_z}\right)}\omega, \bar{\omega}_m|_{in} = a\sqrt{\left(\frac{\rho}{G_z}\right)}\omega$ is shown in Tables 7 to 9 and Tables 11 to 13 for thick annular plate with laminate transversely isotropic material for thickness–radius ratios (h/a) 0.3 and circular plate with laminate transversely isotropic material for various thickness–radius ratios (h/a) 0.3 , inner-outer radius ratios (b/a) 0.4 for F–F, F–Ss, F–Sh, F–C,Ss–F, Ss–Ss, Ss–Sh, Ss–C, Sh–F, Sh–Ss, Sh–Sh, Sh–C boundry condition and thickness–radius ratios (h/a) 0.2 , inner-outer radius ratios (b/a) 0.4 for C–F, C–Ss, C–Sh, and C–C boundry condition is shown in tables 10,14 . The symbolism F–Ss indicates that the edges r = Ri and Ro are free and soft simply supported, respectively. The symbolism F–C indicates that the edges r = Ri and Ro are free and clamped, respectively.

## Conclusions:

In this paper, governing equations for freely vibrating circular-annular plates was derived by applying Hamilton's principle, according to the Reddy's third-order shear deformation plate theory) TSDT). Exact closed-form solutions were given to analyze free in-plane and out-plane vibration of circular-annular plates under different combinations of free, soft simply supported, hard simply supported and clamped boundary conditions. In-plane and out-plane non-dimensional natural frequencies for various boundary conditions were tabulated based on the TSDT and 3-D FEM .in-plane and out-plane mode shapes were presented to illustrate flexural motions of the plate.

**Appendix A**

The coefficients of Eqs. (19) are given by,

$$\overline{\alpha_1} = \sum_{k=1}^{N} \frac{E^k}{1-(v^k)^2}(z_{k+1} - z_k)$$

$$\overline{\alpha_2} = \frac{1}{2}\sum_{k=1}^{N} \frac{E^k}{1+(v^k)}(z_{k+1} - z_k)$$

$$\overline{\alpha_3} = \frac{1}{2}\sum_{k=1}^{N} \frac{E^k}{1-(v^k)^2}\left((z_{k+1}^2 - z_k^2) - \frac{2}{3h^2}(z_{k+1}^4 - z_k^4)\right)$$

$$\overline{\alpha_4} = \frac{1}{4}\sum_{k=1}^{N} \frac{E^k}{1+(v^k)}\left((z_{k+1}^2 - z_k^2) - \frac{2}{3h^2}(z_{k+1}^4 - z_k^4)\right) \quad \text{A-1}$$

$$\overline{\alpha_5} = \frac{1}{3h^2}\sum_{k=1}^{N}\frac{E^k}{1-(\nu^k)^2}(z_{k+1}^{4} - z_k^{4})$$

$$\overline{\alpha_6} = \frac{1}{3}\sum_{k=1}^{N}\frac{E^k}{1-(\nu^k)^2}\left((z_{k+1}^{3} - z_k^{3}) - \frac{8}{5h^2}(z_{k+1}^{5} - z_k^{5}) + \frac{16}{21h^4}(z_{k+1}^{7} - z_k^{7})\right)$$

$$\overline{\alpha_7} = \frac{1}{6}\sum_{k=1}^{N}\frac{E^k}{1+(\nu^k)}\left((z_{k+1}^{3} - z_k^{3}) - \frac{8}{5h^2}(z_{k+1}^{5} - z_k^{5}) + \frac{16}{21h^4}(z_{k+1}^{7} - z_k^{7})\right)$$

$$\overline{\alpha_8} = \frac{1}{6}\sum_{k=1}^{N}G_3^{(k)}\left((z_{k+1} - z_k) - \frac{8}{3h^2}(z_{k+1}^{3} - z_k^{3}) + \frac{16}{5h^4}(z_{k+1}^{5} - z_k^{5})\right)$$

$$\overline{\alpha_9} = \frac{1}{3h^2}\sum_{k=1}^{N}\frac{E^k}{1-(\nu^k)^2}\left(\frac{1}{5}(z_{k+1}^{5} - z_k^{5}) - \frac{4}{21h^2}(z_{k+1}^{7} - z_k^{7})\right)$$

$$\overline{\alpha_{10}} = \frac{16}{63h^4}\sum_{k=1}^{N}\frac{E^k}{1-(\nu^k)^2}(z_{k+1}^{7} - z_k^{7})$$

And,

$$\overline{I_{11}} = \sum_{k=1}^{N}\rho^{(k)}(z_{k+1} - z_k)$$

$$\overline{I_{22}} = \frac{1}{2}\sum_{k=1}^{N}\frac{E^k}{1-(\nu^k)^2}\left((z_{k+1}^{2} - z_k^{2}) - \frac{2}{3h^2}(z_{k+1}^{4} - z_k^{4})\right)$$

$$\overline{I_{33}} = \frac{1}{3}\sum_{k=1}^{N}\frac{E^k}{1-(\nu^k)^2}\left((z_{k+1}^{3} - z_k^{3}) - \frac{8}{5h^2}(z_{k+1}^{5} - z_k^{5}) + \frac{16}{21h^2}(z_{k+1}^{7} - z_k^{7})\right) \qquad \text{A-2}$$

$$\overline{I_{44}} = \frac{1}{3h^2}\sum_{k=1}^{N}\rho^{(k)}(z_{k+1}^{4} - z_k^{4})$$

$$\bar{I}_{55} = \frac{1}{3h^2}\sum_{k=1}^{N} \rho^{(k)}\left(\frac{1}{5}(z_{k+1}{}^5 - z_k{}^5) - \frac{4}{21h^2}(z_{k+1}{}^7 - z_k{}^7)\right)$$

$$\bar{I}_{77} = \frac{16}{63h^4}\sum_{k=1}^{N} \rho^{(k)}(z_{k+1}{}^7 - z_k{}^7)$$

**Appendix B**

The coefficients $\bar{\ell}_i (i = 1,2,…,10)$ are defined as

$$\bar{\ell}_1 = \bar{\alpha}_2$$

$$\bar{\ell}_2 = \bar{\ell}_4 = \bar{I}_{11}\omega^2$$

$$\bar{\ell}_3 = \bar{\alpha}_1$$

$$\bar{\ell}_5 = \bar{\alpha}_7$$

$$\bar{\ell}_6 = \bar{\alpha}_8 - \bar{I}_{33}\omega^2$$

$$\bar{\ell}_7 = \frac{\bar{\alpha}_{10}\bar{\alpha}_6 - \bar{\alpha}_9{}^2}{\bar{\alpha}_8}$$

$$\bar{\ell}_8 = \left(\frac{\bar{I}_{33}\bar{\alpha}_{10} - 2\bar{I}_{55}\bar{\alpha}_9 + \bar{I}_{77}\bar{\alpha}_6}{\bar{\alpha}_8}\right)\omega^2 - (\bar{\alpha}_6 + 2\bar{\alpha}_9 + \bar{\alpha}_{10})$$

$$\bar{\ell}_9 = \left(\frac{\bar{I}_{55}{}^2 - \bar{I}_{33}\bar{I}_{77}}{\bar{\alpha}_8}\omega^2 + \frac{\bar{\alpha}_6}{\bar{\alpha}_8}\bar{I}_{11} + \bar{I}_{33} + 2\bar{I}_{55} + \bar{I}_{77}\right)\omega^2$$